\documentclass[a4paper,10pt]{article}
\usepackage{amsmath,amssymb,mathtools}  
\usepackage{soul}                       
\usepackage[usenames,dvipsnames]{xcolor}
\usepackage[raggedright]{titlesec}
\usepackage{scrextend}
\deffootnote{2em}{0em}{\thefootnotemark\quad}
\usepackage{multicol}
\usepackage[left=0.75in,right=0.75in,top=1.0in,bottom=1.25in]{geometry}
\usepackage{hyperref}
\usepackage{float}
\usepackage{xcolor}

\hypersetup{
  colorlinks   = true, 
  urlcolor     = blue, 
  linkcolor    = blue, 
  citecolor   = blue 
}

\setstcolor{red}    
\setulcolor{red}    
\allowdisplaybreaks 

\let\oldabstract\abstract
\let\oldendabstract\endabstract
\makeatletter
\renewenvironment{abstract}
{%
               {\list{}{\addtolength{\leftmargin}{4em} 
                        \listparindent 1.5em%
                        \itemindent    \listparindent%
                        \rightmargin   \leftmargin%
                        \parsep        \z@ \@plus\p@}%
                \item\relax}%
               {\endlist}%
\oldabstract}
{\oldendabstract}
\makeatother

\title{\LARGE\textbf{\textsf{Percolation of `Civilisation' in a Homogeneous Isotropic Universe}}}

\author{\normalsize Allan L. Alinea, Cedrix Jake C. Jadrin }

\date{}

\begin{document}

\maketitle
\vspace{-2.25em}
\noindent
\begin{center}
{\small Institute of Mathematical Sciences and Physics,\\ University of the Philippines Los Ba\~nos,\\College, Los Ba\~nos, Laguna 4031 Philippines, alalinea@up.edu.ph}
\end{center}

\bigskip
\begin{abstract}
	\noindent{ In this work, we consider the spread of a `civilisation' in an idealised homogeneous isotropic universe where all the planets of interest are habitable. Following a framework that goes beyond the usual idea of percolation, we investigate the behaviour of the number of colonised planets with time, and the total colonisation time for three types of universes. These include static, dark energy-dominated, and matter-dominated universes. For all these types of universes, we find a remarkable fit with the Logistic Growth Function for the number of colonised planets with time. This is in spite of the fact that for the matter- and dark-energy dominated universes, the space itself is expanding. For the total colonisation time, $T$, the case for a dark energy-dominated universe is marked with divergence beyond the linear regime characterised by small values of the Hubble parameter, $H$. Not all planets in a spherical section of this universe can be `colonised' due to the presence of a shrinking Hubble sphere. In other words, the recession speeds of other planets go beyond the speed of light making them impossible to reach. On the other hand, for a matter-dominated universe, while there is an apparent horizon, the Hubble sphere is growing instead of shrinking. This leads to a finite total colonisation time that depends on the Hubble parameter characterising the universe; in particular, we find $T\sim H$ for small $H$ and $T\sim H^2$ for large $H$.}
\bigskip
\\
\noindent
{\small\textbf{keywords:} \textit{percolation, homogeneous isotropic universe, dark energy-dominated, matter-\phantom{dominatedd}dominated, Hubble horizon, FLRW metric}}	
\end{abstract}

\bigskip
\begin{multicols}{2}
\section{Introduction}
\label{secIntro}
\noindent The question of whether we are alone or not in the Universe dates back to antiquity; perhaps, as early as the times when modern humans started wondering about the night sky. With the realisation that the sun is simply one of the seemingly countless stars out there, many of which hosting their own solar systems \cite{Conselice:2016, Lauer:2021, esagaia:2023}, we gained a dazzling insight that the probability of existence of life beyond the Sun's influence, cannot be zero. In fact, counting the number of solar systems and planets in the \textit{Goldilocks zone}, there could be a multitude of planets in our galaxy and beyond, harnessing life \cite{Lorenz:2020,Petigura:2013}. Pushing this idea further, with about a 14-billion year old Universe based on the $\Lambda$CDM model \cite{Dodelson:2021,Planck:2018vyg,Gron:2002}, life in other planets may not be limited to plants and wild animals alone, but could include more technologically advanced civilisations compared to ours.

 In spite of this, we have yet to find a definitive proof for the existence of alien life, much less made a contact with intelligent {extraterrestrial} beings. It is possible that advanced alien civilisations, should they exist, prefer not to interfere with human affairs or {those} of other intelligent beings, in general; letting them take their own course of development \cite{Brin:1983, Annis:1999}. Or, perhaps more reasonably, the density of `living' planets in the Universe is extremely low \cite{Vakoch:2015}. The extremely large distances between these `living' planets and an even more sparse distribution of putative intelligent civilisations, effectively hinder discovery of alien life and communication amongst  intelligent beings in different solar systems. Possibly, it is also for this main reason that the colonisation of other planets and the propagation of civilisations in the network of stars constituting the Universe is extremely slow. Although it is about 14 billion years old, it may {not be} old enough for life to significantly propagate to habitable or terraformable planets.\footnote{Readers interested in the so called \textit{Fermi-paradox} and colonisation, their analysis and/or proposed solutions within the framework of an effectively static section(s) of the Universe and percolation/cellular automata, may refer to Refs. \cite{Landis:1998,Galeraetal:2019,Hairetal:2019,Lingam:2019,Bezsudnov:2010} and the references therein.}

In line with this, one may wonder from a computational physics perspective, given a spherical section of the Universe that could be bounded by a Hubble horizon\footnote{For cosmological perturbations in the early Universe and its relation to the horizon, readers may refer to Refs. \cite{Alinea:2015pza,Alinea:2016qlf,Alinea:2017ncx,Alinea:2020laa}.}\cite{Weinberg:2008zzc}, how long would it take an advanced civilisation to colonise all the habitable or terraformable planets in it? Coupled with this question, given long enough time, can an advanced civilisation possibly colonise all habitable or terraformable planets in a given spherical section of the Universe? Considering the dynamics of colonisation, we may also ask, with the progressive spreading of a civilisation, how would the number of occupied planets behave with time? Given the complexity of the Universe, taking into account the variation in planet characteristics, conditions of host stars, and variations in shapes and age of galaxies, amongst others, these and other related questions can be challenging to fully answer. Nonetheless, if we are to look for the distant future of humanity, noting that we are most likely bound to leave the solar system, they are sensible questions to ponder rooted in our impulse to survive and inclination to know the cosmos.

This study aims to answer the three questions mentioned above{,} albeit within a more modest and doable framework\footnote{Perhaps a good realistic step beyond the idealisation in this work is by using self-replicating probes \cite{Wiley:2015,Borgue:2021} for space exploration.}. The idea is to take small but significant steps by investigating a simplified system to pave the way for more in-depth and complex researches in the future. As part of our simplifying assumptions, we consider a homogeneous and isotropic system, that is, {an} FLRW universe \cite{Dodelson:2021,Weinberg:2008zzc}, composed solely of cells representing habitable planets. Then we take a three-dimensional spherical section of this system and let a `civilisation' start at the centremost cell. From here, it spreads to other cells with a constant propagation speed. The nearest cells are occupied first, and each newly occupied cell becomes a new \textit{undying} source for the propagation of `civilisation' to neighbouring cells; please refer to the following sections for more details. Given this scenario, we investigate three cases, namely, (a) static, (b) dark energy-dominated, and (c) matter-dominated universes \cite{Rubakov:2017xzr}. In contrast to the first one, the last two are both dynamically expanding universes with different expansion rates dictated by {their} matter-energy content.

At a glance, this may look like the propagation of life or civilisation modeled based on \textit{percolation theory} \cite{Giordano:2005,Boudreau:2018,Sahimi:1994,Basta:1994,Paniagua:1997}. Indeed, it is similar to those familiar cases involving the spread of forest fire \cite{Beer:1990} or disease \cite{Mello:2021}, implemented on a square or cubic lattice. However, the current study is interesting in that it differs from the usual `textbook' percolation in at least two ways. First, the cells representing planets are not arranged following a regular lattice\footnote{For percolation on irregular but static lattice, curious readers may refer to Ref. \cite{Ren:2017} and references therein.}. Rather, they are located randomly inside a sphere. However, because the locations are random with each point inside the sphere \textit{equally} likely to have a cell, the system is effectively homogeneous and isotropic. Second, apart from the first case of a static universe, the space itself is expanding\footnote{This is one of the main reasons why it is not straightforward to implement the propagation of `civilisation', as we describe it in this paper, as the usual percolation on a square or cubic lattice found in textbooks on computational physics  \cite{Giordano:2005,Boudreau:2018,Sahimi:1994}.}. The expansion follows that of an FLRW universe (like our own Universe) described by the metric given by \cite{Dodelson:2021,Weinberg:2008zzc}
\begin{align}
   \label{flrwmetric}
   { ds^2 = -dt^2 + a^2(t) \delta_{ij} dx^i dx^j},
\end{align}
where $t$ and $x^i$ are the time and spatial \textit{comoving} coordinates, respectively, $\delta_{ij}$ is the Kronecker delta, and $a(t)$ is the scale factor; here, the Einstein summation convention is followed for the spatial indices $i,j = 1,2,3$. With these two modifications, particularly the second one, our work may be seen as an effective extension of percolation theory to the regime involving an irregular and dynamical lattice. 

With our sight to fulfill the objectives of this work, we divide the rest of the paper into four sections. Section \ref{secstatic} involves the case for a static universe. As we shall see, starting from the centremost cell, `civilisation' propagates to other cells with time following the Logistic Growth Model \cite{Tsoularis:2002}. In Sec. \ref{secdarkE}, we deal with an expanding universe driven by a constant dark energy in Einstein's general theory of relativity. The way of propagation of `civilisation' is the same as that for the static universe but the space where the cells are embedded is expanding. We shall find that the unabated expansion due to dark energy leads to the existence of a horizon preventing the occupation of all cells in our sphere. Following this is another case of expanding universe, but this time, it is driven by matter instead of dark energy. In Sec. \ref{secmatter}, we cover a matter-dominated universe with a `tamer' expansion rate. We shall see that although there appears to be a `horizon' as in Sec. \ref{secdarkE}, it differs in that the corresponding Hubble sphere grows with time faster than the propagation of `civilisation'. This ultimately results in the occupation of all cells despite the expansion of the universe. Lastly, we state our concluding remarks and future prospects in Sec. \ref{seConclude}.

\section{Static Universe}
\label{secstatic}

\begin{figure}[H]
    \centering
    \includegraphics[scale=0.70]{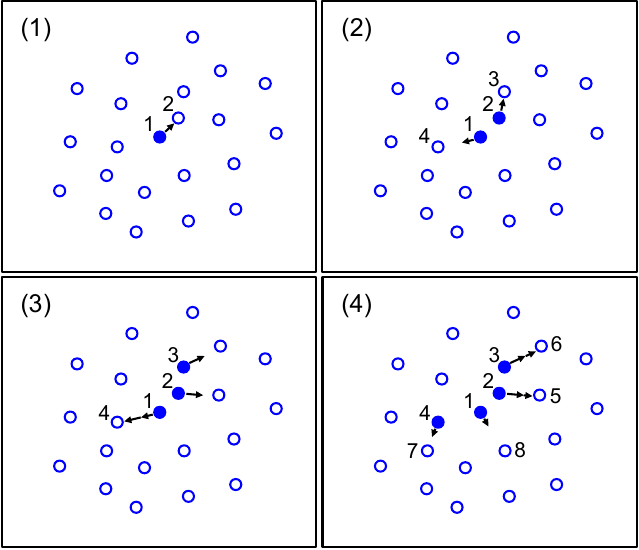} 
    \caption{Simplified illustration for the occupation of cells or `planets' in a static universe. The centremost cell is occupied first and `life' or `civilisation' propagates towards neighbouring cells following the order (1) $\rightarrow$ (2) $\rightarrow $ (3) $\rightarrow$ (4).}
    \label{figsimpstat}
\end{figure}

\noindent Following a short description about the propagation of civilisation in the preceding section, we consider an \textit{ideal} static universe whose only constituents are habitable planets represented by cells in an empty {three dimensional} space. {Considering a spherical section of this universe, we have the following general algorithm for our simulation:}
\begin{itemize}
	
	\small
	\item 
	Randomly distribute $N$ number of cells inside the spherical section following a \textit{uniform} probability distribution corresponding to a homogenenous isotropic universe. In our program, this is done using Permuted Congruential Generator \cite{Oneill:2014} for the cell coordinates $(x,y,z)$ with the constraint $x^2 + y^2 + z^2 < r^2$, where $r$ is the radius of the spherical section.
	\item 
	Initially, the centremost cell represent the only `living' planet; all other cells are yet to be occupied. The time counter for the spread of civilisation is set to zero and the counter for the number of occupied cells is set to one. The civilisation from this cell then propagates following a simplified 2D pattern shown in \textit{Fig.} {\ref{figsimpstat}}; the actual simulation involves a 3D space.
	\item 
	From the centremost cell (1), the civilisation travels at constant speed, $v$, to the nearest cell (2), where a new civilisation is established; ideally, within effectively zero amount of time upon arrival as in the usual percolation of forest fire in a rectangular lattice. 
	\item 
	With two cells (1 and 2) now occupied in the figure, these two civilisations propagate to their corresponding nearest target cells (3 and 4, respectively). The main idea is, whenever a cell is occupied, its civilisation spreads to the nearest cell at the same speed $v$.
	\item 
	Programmatically, each cell has several flags and state variables indicating whether they are (a) inhabited or not, (b) targeted or not by a neighbouring `civilised' cell, and (c) measuring the time of travel to a neighbouring cell. Each time a new `uncivilised' cell is `civilised', the time counter and number of occupied cells are incremented based on (c) and one, respectively. The cells are stored in an array of structures with these flags.
	\item 
	Because neigbouring cells have different distances from the cells targeting them, they are not reached at the same time. In the figure, cell 2 reaches its target cell 3 first before cell 1 reaches its target cell 4 because the distance between cells 2 and 3 is smaller than that for cells 1 and 4. The propagation of civilisation continues from here following the same logic until all cells are occupied or `civilised'.
\end{itemize}

\noindent

Note that in our model, all occupied cells are \textit{undying}. Once a cell is occupied, it remains alive and a source of spreading civilisation throughout the duration of the simulation\footnote{We leave it for future work to deal with death of a civilisation, variation in planet habitability, limited propagation, etc.}. Furthermore, each occupied cell only targets one cell at a time for the propagation of its civilisation---the nearest unoccupied cell. Given the highly unlikely scenario where two neighbouring nearest cells are of (nearly) equal distance from an occupied cell, only one of  these two becomes a target; the target is chosen randomly. For the time of arrival from one occupied cell located at a position represented by a vector, $\vec r_i$, measured from the centremost cell, to the nearest unoccupied (target) cell at $\vec r_j$, we have $|\vec r_j - \vec r_i|/v$. Certainly, because $v$ is constant, two or more neighbouring occupied cells will, in general, not arrive simultaneously at their target cells as illustrated in \textit{Fig.} \ref{figsimpstat}. This serves to remind us that unlike that of the usual percolation on 2D square lattice (modeling, for instance, the spread of forest fire,) our cells are not on a regular lattice. This leads to non-uniform time steps between occupations of neighbouring cells.

\begin{figure}[H]
    \centering
    \includegraphics[scale=0.70]{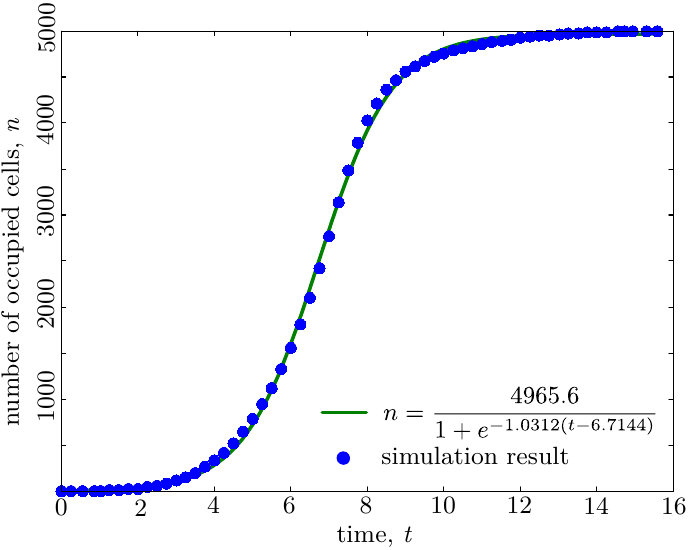}
    \caption{Behaviour of the number of cells occupied with time in a static universe, for $N = 5000$ cells and 500 trials.}
    \label{fignvststat}
\end{figure}

Having established our algorithm above, we perform a simulation involving $N = 5000$ cells located in a sphere with a radius of $L = 5.0$ units. For simplicity, we set the scale factor in the metric given by (\ref{flrwmetric}) and the constant speed of propagation to unity {$(v = 1)$}{. Not to be confused with the speed of light, $c = 1$, usually employed in General Relativity and Quantum Field Theory, the choice of units, $v=1$, simply} implies our unit of distance is the same as the unit of time{; e.g., $1\,\text{km} = 1\,\text{s}$.} \textit{Figure} \ref{fignvststat} shows our simulation result for the number of cells occupied with respect to time. This is an averaged result over 500 trials. For each trial, 5000 cells are randomly distributed inside the sphere using the Permuted Congruential Generator \cite{Oneill:2014} and civilisation is allowed to spread from the centremost cell with the intention to occupy all cells. With 500 trials, the standard errors of the mean times to reach cells 2 to 5000, divided by the corresponding mean times, are all below 0.5\%; e.g., the average time to reach cell 5000 is $\bar t = 15.58$ units with a standard error of $\sigma_{\bar t} = 0.019$ corresponding to $\sigma_{\bar t}/\bar t \approx 0.12\%$. In other words, for the purposes of this study, 500 trials seem to be more than good enough for the `smooth' plot shown in \textit{Fig.} \ref{fignvststat}.

\noindent

Focusing on the behaviour of the spread of civilisation shown in the figure, we find a slow start. With a few sources of civilisations, the spread is correspondingly slow. Then it picks up speed as more and more cells are occupied corresponding to many sources of civilisations. As is evident in the figure, the maximum rate of propagation of civilisation happens when around half of all cells are occupied; about $n = 2500$ corresponding to the approximate time $t \approx 7$ units. Beyond this, while there are more sources of civilisations, there are fewer cells for them to spread to in our sphere of interest. The graph of the number of occupied cells, $n$, tapers off with $t$ near $n = N = 5000$. {Note that increasing the speed, $v$, for the propagation of civilisation can only increase the number of occupied cells with time. But the behaviour of $n$ with $t$ remains the same. Informally, the `S'-shape curve in \textit{Fig.} \ref{fignvststat} is simply `compressed' to the left with higher $v$.}

Knowing this, our ideal model for the spread of civilisation in a static universe, appears to follow the Logistic Growth Model \cite{Tsoularis:2002}. It seems to stand side-by-side with the other applications of this model such as population growth \cite{Tsoularis:2001}, spread of communicable disease \cite{Cheng:2020}, and chemical reactions in a closed system \cite{Yin:2018}, amongst others. The prominent behaviour of these dynamical systems is a relatively slow start in the propagation due to limited sources (e.g., disease carriers). Then, as the number of sources grows, with many others yet to be reached by these sources, the propagation speeds up. This happens until there are significantly more sources than those that can be infected or occupied in cases of disease spread and colonisation, respectively. Eventually, the propagation slows down with less number of cells to occupy (colonisation) or less reactants to react (chemical reactions) until the elements of the system are exhausted.

Indeed, the behaviour of $n$ with respect to $t$ most likely follows that of the Logistic Growth Model. Curve fitting our simulation result with the logistic function given by
\begin{align}
	{n(t) = \frac{N_0}{1 + e^{-k(t - t_0)}}},
\end{align}
where $N_0,\,k,$ and $t_0$ are the adjustable parameters, yields the equation for the best fit curve indicated in \textit{Fig.} \ref{fignvststat}. The corresponding coefficient of determination, $R^2 = 0.9996$, is extremely close to unity, indicating an excellent fit. 

In the next two sections, we shall see if the Logistic Growth Model is still followed when the space itself is expanding; that is, homogeneously and isotropically for dark energy-dominated and matter-dominated universes.

\section{Dark Energy-Dominated Universe}
\label{secdarkE}
\noindent Let us consider a dark energy-dominated universe characterised by a constant Hubble parameter, $H \equiv \dot{a}/a$, where the dot indicates derivative with respect to time. Such a universe corresponds to an exponentially expanding spacetime driven by a constant dark energy, $\Lambda$, in Einstein's general theory of relativity \cite{Dodelson:2021,Weinberg:2008zzc}. Indeed, with $H = \text{const.}$, we see that 
\begin{align}
\label{expscale}
{\frac{da}{a} = H\,dt
\quad\Rightarrow\quad
a\sim e^{Ht}}.
\end{align}
Situating (idealised) habitable planets or cells in this universe for colonisation, as in the immediately preceding section, implies that the `physical' distances between the cells increase with time, even in the absence of individual random motion. It is akin to two dots on the surface of a rubber balloon `moving' farther apart as the balloon is inflated. As such, the time it would take for one civilisation to reach a nearby cell and colonise it, is larger compared to that in a static universe, given the same propagation speed, $v$.

In particular, let us consider two cells located at $\vec r_i$ and $\vec r_j$, respectively, measured from the centremost cell. Noting that our space is coordinatised by comoving coordinates indicated in (\ref{flrwmetric}), the time, $t_j$, it takes to reach cell $j$ from cell $i$ starting at time, $t_i$, is given by 
\begin{align}
	{|\vec r_j - \vec r_i| =
	\int _{t_i}^{t_j} dt \ \frac{v}{a}}.
\end{align}
Needless to say, if $a $ is unity, it reduces to that of the case for static universe in the immediately preceding section; i.e., $|\vec r_j - \vec r_i| = v(t_j - t_i)$. Using the relation for the scale factor given by (\ref{expscale}) above, we find upon integration,
\begin{align}
    \label{tjHrj}
	{t_j
	=
	-\frac{1}{H} 
	\ln\left(
		e^{-H t_i}
		-
		\frac{H}{v} |\vec r_j - \vec r_i|
	\right)}.
\end{align}
If both $|\vec r_j - \vec r_i|$ and $v$ are set to unity, for instance, then for\footnote{We are omitting here explicit mention of units for brevity. As in the previous section, $v$ is set to unity, leading to equality in distance and time units.} $H = 0.1$ and $t_i = 0$, we get $t_j \approx 1.05$ as opposed to the case of static universe where $t_j = 1.00$.

\noindent
\begin{figure}[H]
    \centering
    \includegraphics[scale=0.70]{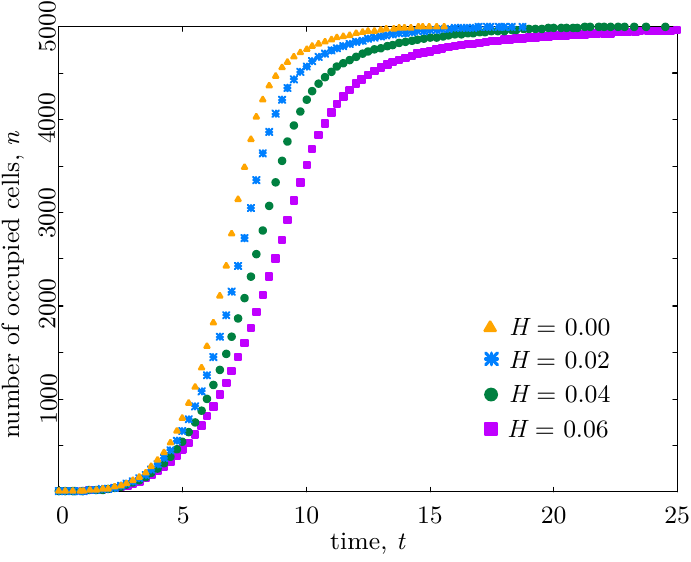}
    \caption{Behaviour of the number of cells occupied with time in a dark energy-dominated universe, for varying values of the Hubble parameter; $N = 5000$ cells and 500 trials.}
    \label{fignvstde0p00to0p06}
\end{figure}

\begin{figure}[H]
    \centering
    \includegraphics[scale=0.70]{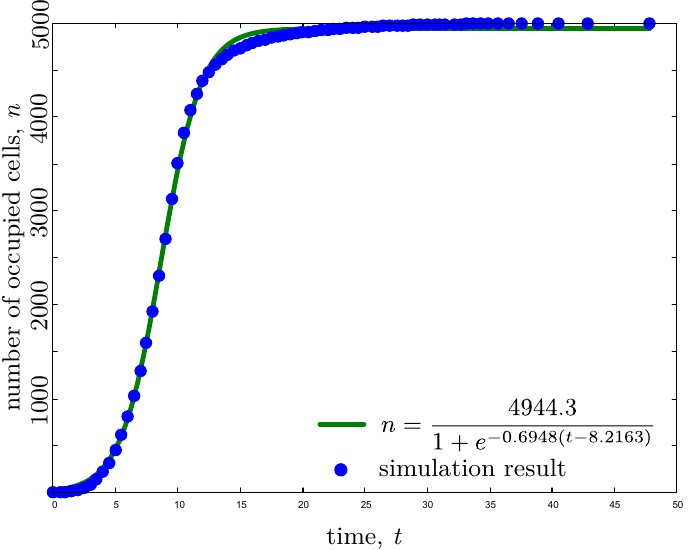}
    \caption{Behaviour of the number of cells occupied with time in a dark energy-dominated universe, for $H = 0.060,\,N = 5000$ cells, and 500 trials.}
    \label{fignvstde0p06}
\end{figure}

\setcounter{footnote}{0}
\textit{Figure} \ref{fignvstde0p00to0p06} shows our simulation result for the number of colonised cells with time in a dark energy-dominated universe characterised by a constant Hubble parameter.  Here, $N = 5000$ cells and the number of trials is 500 for each series\footnote{As in the case for static universe, with 500 trials, the standard errors of the mean times to reach cells 2 to 5000, divided by the corresponding mean times, are all below 0.5\%; e.g., the average time to reach cell 5000 is $\bar t = 30.36$ units with a standard error of $\sigma_{\bar t} = 0.089$ corresponding to $\sigma_{\bar t}/\bar t \approx 0.29\%$.}, as in Sec. \ref{secstatic}. With reference to the static universe ($H = 0.00$), we see a deviation to the right of the number of colonised cells with time. In other words, it takes longer time to colonise a given number of cells from the centremost cell due to the expansion of the Universe, and this time increases with increasing $H$. Nonetheless, looking at the figure, the behaviour of $n$ with $t$ seems to still follow the Logistic Growth Model as in the static universe. In fact, fitting this function to the data points, we find that the coefficients of determination remain extremely closed to unity as we vary $H$ above from 0.00 to 0.06. For $H = 0.06$ (see \textit{Fig.} \ref{fignvstde0p06}), where the deviation is expected to be the highest among the three (non-static universe) series in \textit{Fig.} \ref{fignvstde0p00to0p06}, we find $R^2 = 0.9992 \approx 1$, suggesting an excellent fit.

\noindent

\noindent
\begin{figure}[H]
    \centering
    \includegraphics[scale=0.70]{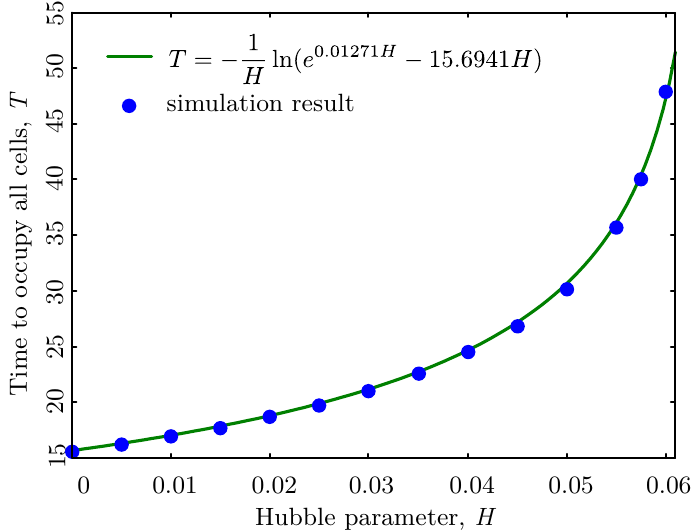}
    \caption{Behaviour of time to occupy \textit{all} cells with respect to the Hubble parameter, in a dark energy-dominated universe, for $N = 5000$ cells and 500 trials.}
    \label{figTvsHde}
\end{figure}

Beyond $H = 0.06$, however, we see an effective breakdown from the Logistic Growth Model. This is because for larger values of $H$, all cells in our sphere can no longer be colonised. This behaviour is apparent in \textit{Fig.} \ref{figTvsHde} showing the time to occupy \textit{all} cells, $T$, with respect to the Hubble parameter\footnote{To make sense of the data here on a galactic scale, consider a sphere the size of the Milky Way galaxy---radius of about $5\times 10^{20}\,\text m$---with only 5000 habitable planets as in our simulation. For the expansion rate of $H = 0.01$ with $v$ corresponding to one percent the speed light, the total colonisation time of 17 units in the figure simply converts to about 17 million years; \textit{i.e.}, 1 time unit $\approx$ 1 million years.}. As the expansion rate increases, $T$ correspondingly increases; that is, in a seemingly linear manner for small $H$ but goes up fast beyond the linear regime. Above $H=0.06$, there seems to be an asymptote where $T$ diverges. Our curve fitting function for $T$ with parameters $t_0$ and $d$, borrowed from (\ref{tjHrj}), captures this behaviour with $R^2 = 0.9990$:

\begin{align}
    \label{fitdarkeT}
    {T = -\frac{1}{H}\ln\big(
        e^{-Ht_0} - Hd
    \big)}.
\end{align}
For $H \approx 0$, its Taylor series expansion yields $T \sim H$ confirming the mentioned linear behaviour. Beyond this is an asymptote given by $e^{-Ht_0} - Hd$. For the results in \textit{Fig.} \ref{figTvsHde}, we have an effective asymptote at $H = 0.06367$.

The existence of a horizon is apparent in (\ref{tjHrj}) describing the time, $t_j$, it takes to reach cell $j$ from cell $i$ starting at time, $t_i$. Given a fixed value of $H$, we find $t_j \rightarrow \infty$ as the argument of the (natural) logarithmic function goes to zero; that is,
\begin{align}
    {e^{-Ht_i} 
    \rightarrow 
    \frac{H}{v}|\vec r_j - \vec r_i|}.
\end{align}
Early on, civilisation can propagate from the centremost cell to the neighbouring cells because the inter-cell physical distances are still small. But as time goes by, the universe expands, and these inter-cell distances increase. Colonisation of new cells takes longer and longer until such time that the next neighbouring cell to be colonised goes beyond the horizon; it requires a propagation velocity higher than $v$ to be reached. 

Certainly, one may consider a greater propagation velocity for the civilisation to spread to all cells. Following this line of thinking, however, will only increase the value of $H$ in \textit{Fig.} \ref{figTvsHde} at which the total colonisation time diverges. To see this, we recall the idea of a Hubble horizon \cite{Dodelson:2021,Weinberg:2008zzc} corresponding to a Hubble sphere with comoving radius defined as
\begin{align}
    {\label{hubblehorizon}
    R_H \equiv \frac{c}{aH}},
\end{align}

\noindent where $c$ is the maximal \textit{causal} velocity that is physically attainable; i.e., the velocity of light. In accord with Hubble's law, the effective recession  velocity of the objects situated on the Hubble sphere relative to its centre is $c$. It follows that the total colonisation time, even in our idealised scenario where the spread of civilisation is made easy by simple propagation to neighbouring cells, is bound to diverge, either with increasing $H$ or with increasing radius of our chosen spherical section of a dark energy-dominated universe. Furthermore, because of this divergence, the breakdown from the Logistic Growth Model for the number of occupied cells with time seems inevitable.

\section{Matter-Dominated Universe}
\label{secmatter}

\noindent For a matter-dominated universe, the Hubble parameter is related to the scale factor as $H^2 \propto a^{-3}$, by virtue of the Friedmann equation \cite{Dodelson:2021,Weinberg:2008zzc,Uzan:2001}. With the definition given by $H \equiv \dot a /a$, we find $a \propto t^{\frac{2}{3}}$.
Following the same logic as in the immediately preceding section, given this relationship between $a$ and $t$, we obtain the time to reach cell $j$ from cell $i$ given by
\begin{align}
\label{tjmatter}
{t_j
=
\frac{2}{3H_0} \bigg[
    (1 + \frac{3}{2} H_0 t_i)^{\frac{1}{3}}	
    +
    \frac{H_0}{2v}
    |\vec r_j - \vec r_i|
\bigg]^{3} 
- 
\frac{2}{3H_0}},
\end{align}
where $H_0$ is the initial Hubble parameter. Needless to say, $H$ is no longer constant unlike that of a dark energy-dominated universe. In our simulation, $H_0$ is chosen to coincide with the corresponding constant $H$ for a dark energy-dominated universe, for the purpose of comparison.

\begin{figure}[H]
    \centering
    \includegraphics[scale=0.70]{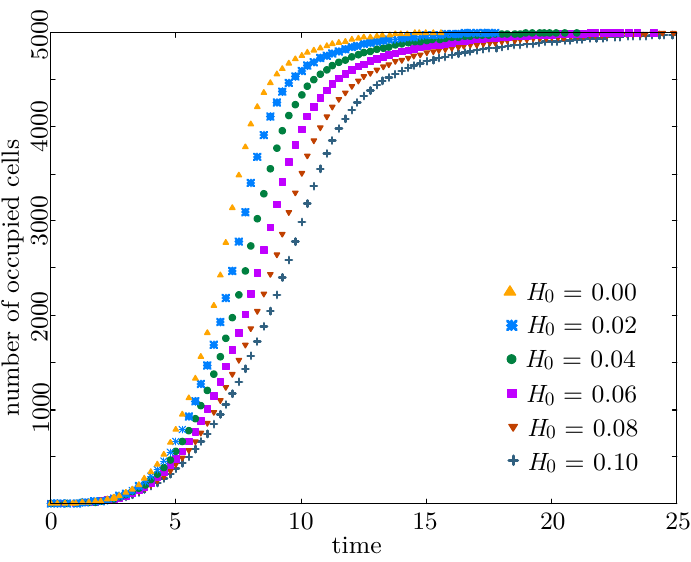}
    \caption{Behaviour of the number of cells occupied with time in a matter-dominated universe, for varying values of the \textit{initial} Hubble parameter; $N = 5000$ cells and 500 trials.}
    \label{fignvstma}
\end{figure}

\textit{Figure} \ref{fignvstma} shows our simulation results for the number of occupied cells with time for different values of $H_0$. Similar to that for the dark energy-dominated universe, the time it takes to occupy a given number of cells is higher for higher initial value of the Hubble parameter. Furthermore, a short visual inspection tells us that the graphs seem to follow the same pattern; that is, a logistic growth function. Indeed, the coefficients of determination in the curve fitting using this function, remain very close to unity for all the series in the figure; for the highest $H_0$ in the graph, $R^2 = 0.9994$, indicating an excellent fit. Having said this, we notice that on average, the time it takes to occupy a certain number of cells is lower compared to that of the dark energy-dominated universe. For instance, the times to reach 4500 cells for $H = 0.06$ are 11.5 and 12.6 (time units) for the matter- and dark energy-dominated universes, respectively. This is justified by the fact that although both universes start with the same value of the Hubble parameter, that of the dark energy-dominated universe is constant while the other one decreases with time; in particular, for the latter, $H \sim 1/t$. Physically, beyond the initially set time and Hubble parameter, colonisation is faster for a matter-dominated universe because it expands slower compared to a dark energy-dominated universe.

\noindent

\begin{figure}[H]
    \centering
    \includegraphics[scale=0.70]{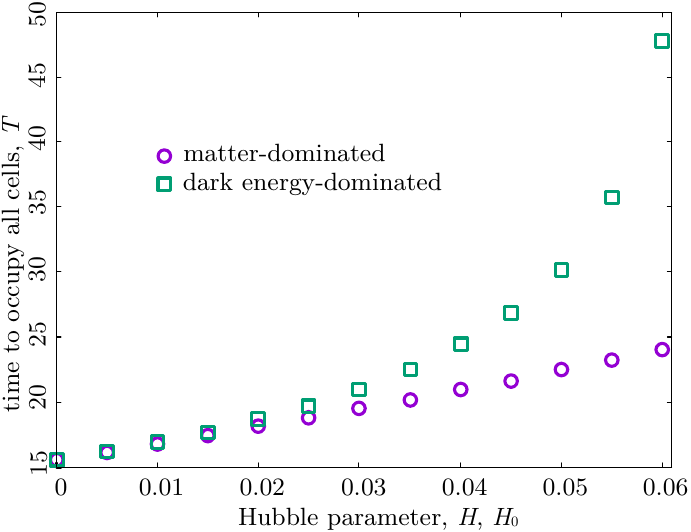}
    \caption{Comparison of the behaviour of the total times to occupy all cells with respect to the (initial) Hubble parameter, in a matter-dominated universe and dark energy dominated universe, for $N = 5000$ cells and 500 trials.}
    \label{figTvsHdema}
\end{figure}

\begin{figure}[H]
    \centering
    \includegraphics[scale=0.70]{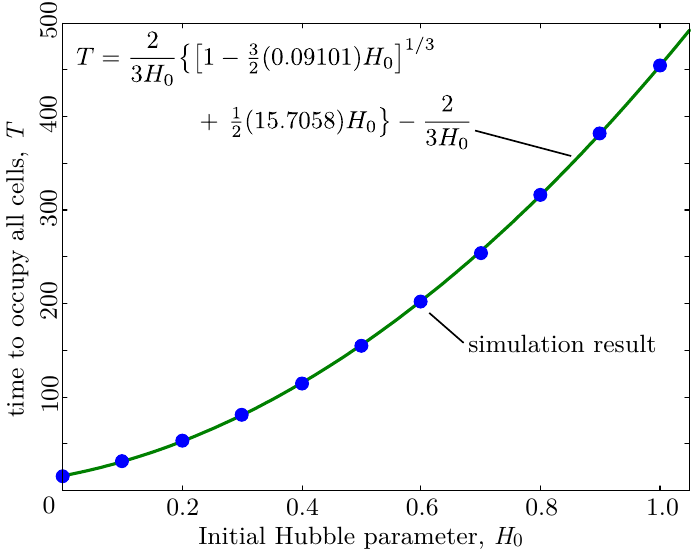}
    \caption{Behaviour of the time to occupy all cells with respect to the Hubble parameter, in a matter-dominated universe, for $N = 5000$ cells and 500 trials.}
    \label{figTvsHma}
\end{figure}

Consistent with this observation, the time to occupy \textit{all} cells in our simulation is correspondingly lower for a matter-dominated universe; see \textit{Fig.} \ref{figTvsHdema}. Furthermore, the figure shows that while $T$ for a dark energy-dominated universe grows fast in the region $0.0 < H \le 0.06$ and even `blows up' around $H = 0.06$, for a matter-dominated universe, $T$ seems to follow a linear behaviour with $H_0$. In fact, as can be seen in \textit{Fig.} \ref{figTvsHma}, it crosses the `dangerous' mark about $H = 0.06$ for the dark energy-dominated universe without problem; that is, as far as $H_0 = 1.0$ in our simulation. Our curve fitting function with parameters $t_0$ and $d$, borrowed from (\ref{tjmatter}), captures the behaviour of $T$ with $H_0$ as

\begin{align}
\label{fitmaxTmatter}
{T
=
\frac{2}{3H_0} \bigg[
    (1 + \frac{3}{2} H_0 t_0)^{\frac{1}{3} }	
    +
    \frac{H_0 d}{2v}
\bigg]^{3} 
- 
\frac{2}{3H_0}}.
\end{align}

\noindent For the result shown in \textit{Fig.} \ref{figTvsHma}, we have $R^2 = 1.0000$, signifying an excellent fit. In agreement with our simulation result and the aforementioned observations, for small values of $H_0$, the Taylor series expansion of (\ref{fitmaxTmatter}) tells us that $T \sim H_0$. Furthermore, for large values of $H_0$, the time to colonise all cells scales as $H_0^2$; that is, a dominantly quadratic behaviour. Lastly, in contrast to (\ref{fitdarkeT}) for a dark energy-dominated universe, the relationship given by (\ref{fitmaxTmatter}) for a matter-dominated universe, has no asymptote marking the divergence of $T$.

\noindent
\noindent
In spite of these observations, one may think more deeply if the time for a civilisation to propagate to \textit{all} cells can remain finite given a finite number of cells but with large finite Hubble parameter. In hindsight, following the same logic as in the previous section, the fact that a matter-dominated universe also expands homogeneously and isotropically seems to point to the existence of a Hubble horizon. Beyond the Hubble sphere, the cells recede from the center at a velocity larger than that of light in accord with Hubble's law. Equation (\ref{hubblehorizon}) for the Hubble horizon, applied for a matter-dominated universe, seems to indicate an apparent horizon that could give rise to a divergent $T$; indeed, a short inspection of (\ref{hubblehorizon}) tells us that $R_H \ne \infty$. However, we cut short this line of thinking for a divergent $T$ by recalling that for a matter-dominated universe, $a \propto t^{\frac{2}{3}}$. This certainly leads to a finite $R_H \sim t^{\frac{1}{3}}$ but the Hubble sphere is \textit{growing}. This is in contrast to the case of dark energy-dominated universe where the comoving Hubble sphere is \textit{shrinking}. Now, multiplying $R_H$ by $a$ to get the `proper' Hubble radius and taking the time derivative of the result yields $c = (aR_H)\dot{}$. We find that although there is an apparent Hubble horizon, the expansion rate of the Hubble sphere relative to its centre is equal to $c$! Consequently, given enough time, cells initially located beyond the horizon eventually enter the Hubble sphere enabling effectively all cells in a spherical section of a matter-dominated universe to be colonised.

\section{Concluding Remarks}
\label{seConclude}

\noindent In the quest to understand the colonisation potential of an advanced civilisation across the vast expanse of habitable planets in our universe, we ponder in this work interesting questions from the perspective of computational physics. These questions involve the behaviour of the number of planets colonised with time, and the total colonisation time with respect to the Hubble parameter describing the universe. Inspired by Percolation Theory, we investigate these questions within an idealised scenario {involving}  homogeneous isotropic universe within which habitable planets are embedded. We study three types of universes, namely, static, dark energy-dominated, and matter-dominated.

We find that the growth in the number of colonised planets with time in a spherical section of a static universe follows the Logistic Growth Model. Such a behaviour confirms our expectation based on its other applications such as the propagation of fire, spread of disease, and chemical reaction. In contrast, the same might not be anticipated for a dark energy- and matter-dominated universes because they are characterised by an expanding space. Surprisingly, our results indicate that even with this expansion, the behaviour of the number of colonised planets with time, fits well with the logistic growth function, albeit with slower occupation or colonisation rate and a caveat for a dark energy-dominated universe.

The case for a dark energy-dominated universe is characterised by a divergent total colonisation time. While it behaves linearly for low values of the  Hubble parameter describing the universe, beyond some cut-off, it blows up marking a breakdown from the Logistic Growth Model. The underlying reason is the existence of a Hubble horizon corresponding to a shrinking Hubble sphere; planets beyond this `moves' faster than the speed of light preventing {further} colonisation. For a matter-dominated universe, the total colonisation time behaves linearly {for} small values of the Hubble parameter and quadratically for large values of this parameter. Our simulation results suggest that the total colonisation for a spherical section of this universe remains finite for arbitrarily large values of the initial Hubble parameter. This is in spite of the existence of an apparent Hubble horizon. We reason that while there is a finite Hubble horizon for a matter-dominated universe, it is growing instead of shrinking as in the case of a dark energy dominated universe. Since its growth is faster than the rate of colonisation, the colonisation of all planets in a spherical section of this universe seems to be always possible.

{In spite of} the limited ideal framework of this study, we find interesting results about colonisation of planets in the universe from the perspective of computational physics. However, our humble beginning leaves a lot of avenues for future exploration. This may include factors involving (a) {planet habitability}, (b) death or survival rate {of civilisations}, and (c) multiple starting civilisations, amongst others. {The first two are certainly related and the addition of the third one requires more programming resources. Focusing on the death rate alone for simplicity, we speculate that for a constant probability of death upon arrival to previously uninhabited planets, the first order behaviour, $T\sim H$, for small $H$, would remain the same, albeit $n$ would grow slower with $t$, for both matter- and dark-energy dominated universes\footnote{If the constant death rate corresponds to effectively lowering the propagation velocity of civilisation in the curve fitting functions for $T$ with respect to $H$ for both matter- and dark energy-dominated universes, then the first order behaviour, $T\sim H$, for small $H$, remains the same based on (\ref{fitdarkeT}) and (\ref{fitmaxTmatter}).}. Spherical sections of both static and matter-dominated universes would continue to be fully occupied given enough time. However, for a dark-energy dominated universe, the breakdown from the Logistic Growth Model for $n$ with respect to $t$ would be earlier given the slower effective propagation rate of civilisation with a shrinking Hubble sphere.} Holding on tight with our effective extension of Percolation Theory, we hope to gain deeper insight with the inclusion of these factors in the study of the spread of civilisation in a homogeneously and isotropically expanding Universe.
\section*{Acknowledgement}
\label{acknow}
\noindent CJC Jadrin would like to express his sincere gratitude to the Department of Science and Technology Science Education Institute (DOST-SEI) for the financial support during the conduct of this study.

\end{multicols}
\end{document}